\documentclass[fleqn,10pt,twocolumn]{SICE_FES26}

\usepackage{amsmath,amsfonts,amssymb}
\usepackage{mathtools}
\usepackage{bm}
\usepackage{algorithm}
\usepackage[noend]{algpseudocode}
\usepackage{physics}

\newcommand{\bbN}{{\mathbb{N}}}
\newcommand{\bbZ}{{\mathbb{Z}}}
\newcommand{\bbR}{{\mathbb{R}}}

\newcommand{\bmx}{{\bm{x}}}
\newcommand{\bmm}{{\bm{m}}}
\newcommand{\bmc}{{\bm{c}}}
\newcommand{\bme}{{\bm{e}}}
\newcommand{\bmk}{{\bm{k}}}

\newcommand{\calI}{{\mathcal{I}}}

\newcommand{\crs}{{\mathsf{crs}}}
\newcommand{\pk}{{\mathsf{pk}}}
\newcommand{\sk}{{\mathsf{sk}}}

\newcommand{\Setup}{{\mathsf{Setup}}}
\newcommand{\KeyGen}{{\mathsf{KeyGen}}}
\newcommand{\Enc}{{\mathsf{Enc}}}
\newcommand{\Dec}{{\mathsf{Dec}}}
\newcommand{\Add}{{\mathsf{Add}}}
\newcommand{\pMult}{{\mathsf{pMult}}}
\newcommand{\Mult}{{\mathsf{Mult}}}
\newcommand{\Pack}{{\mathsf{Pack}}}
\newcommand{\Unpack}{{\mathsf{Unpack}}}
\newcommand{\Rotate}{{\mathsf{Rotate}}}
\newcommand{\num}{{\mathsf{num}}}
\newcommand{\den}{{\mathsf{den}}}

\DeclarePairedDelimiter{\floor}{\lfloor}{\rfloor}
\DeclarePairedDelimiter{\ceil}{\lceil}{\rceil}
\DeclarePairedDelimiter{\round}{\lfloor}{\rceil}

\algrenewcommand\algorithmicindent{1em}

\algrenewcommand\algorithmicdo{}

\newcommand{\figref}[1]{Fig.~\ref{#1}}
\newcommand{\tabref}[1]{Table~\ref{#1}}
\renewcommand{\algref}[1]{Algorithm~\ref{#1}}

\title{Encrypted Visual Feedback Control Using RLWE-Based Cryptosystem}

\author{Taichi Ikezaki${}^{1\dagger}$ and Kaoru Teranishi${}^{2}$}
\speaker{Taichi Ikezaki}

\affils{${}^{1}$Faculty of Environmental, Life, Natural Science and Technology, Okayama University, Okayama, Japan\\
(Tel: +81-08-6251-8644; E-mail: t.ikezaki@okayama-u.ac.jp)\\
${}^{2}$Department of Information and Physical Sciences, Graduate School of Information Science and Technology,\\
The University of Osaka, Osaka, Japan\\
(Tel: +81-06-6879-4247; E-mail: k-teranishi@ist.osaka-u.ac.jp)\\
}
\abstract{%
This study proposes an encrypted visual feedback control algorithm for regulating a one-dimensional stage using Ring Learning With Errors (RLWE) encryption.
The proposed algorithm performs both feature extraction and controller computations directly on encrypted images, ensuring that sensitive visual data remain protected throughout the entire control process.
Furthermore, an image captured by the camera is encrypted into a single ciphertext leveraging the message packing technique of RLWE encryption, thereby reducing computational cost.
The effectiveness of the proposed framework is demonstrated through numerical simulations.
}


\begin{document}

\maketitle

\section{Introduction}

Visual feedback control is a method that utilizes real-time image data from cameras to control dynamical systems~\cite{francois06,francois07}.
By mimicking human visual perception, a visual feedback controller extracts geometric or photometric features from the captured images and computes the corresponding control actions.
This approach enables advanced motion planning and precise manipulation, making it applicable to a wide range of fields, such as autonomous driving~\cite{Hongliang24} and robotics~\cite{zhou26}.

Despite its versatility, the use of images in feedback control introduces significant security and privacy concerns, especially in networked control systems.
Visual data are inherently rich in information, often containing sensitive details about the operational environment, proprietary products, or personal privacy.
In such systems, the captured images are transmitted to a server (e.g., a commercial cloud platform) via public networks, where the controller computation is performed.
Hence, if the communication channels are intercepted by adversaries, or if the server is untrusted, the compromise of security and privacy becomes unavoidable.

To mitigate these risks, encrypted control has emerged as a promising countermeasure~\cite{Kogiso15}.
By leveraging homomorphic encryption, this approach allows a server to compute control actions directly over encrypted measurements and parameters without decryption.
This paradigm ensures security and privacy throughout the control computation, from sensor outputs to actuator inputs.
While encrypted control has been successfully applied to a variety of control methods~\cite{Darup21,Kim22,Schluter23}, its application to visual feedback control remains challenging.
This difficulty primarily stems from the prohibitive computational cost of processing visual images over their encrypted data.

In this paper, we propose an encrypted visual feedback control algorithm tailored for the positioning of a one-dimensional stage using Ring Learning with Errors (RLWE)-based encryption.
The proposed algorithm extracts the geometric centroid of the stage from an encrypted image and computes an encrypted control input to regulate the stage based on this extracted feature.
To overcome the aforementioned computational bottleneck, we apply the message packing technique supported by RLWE encryption schemes when encrypting the images.
To the best of our knowledge, the proposed algorithm is the first realization of encrypted visual feedback control using message packing.
This approach bridges the gap between secure computation and visual feedback control by reducing the computational costs imposed by encryption.

The remainder of this paper is organized as follows.
Section~2 provides an overview of RLWE-based encryption and the message packing technique.
Section~3 introduces the visual feedback control system considered in this paper and presents our proposed encrypted control algorithm.
Section~4 verifies the feasibility of the proposed algorithm through a numerical simulation.
Finally, Section~5 concludes the paper and presents future work.

\section{RLWE encryption}

This section provides a brief overview of RLWE encryption used in this study.
Let $R \coloneqq \bbZ[X] / (X^N + 1)$ be the polynomial ring consisting of polynomials of degree less than $N$ with integer coefficients.
The corresponding addition and multiplication are defined as standard polynomial addition and multiplication modulo $X^N + 1$, i.e., $X^N$ is replaced by $-1$.
Let $\bbZ_q \coloneqq \bbZ \cap [-q/2, q/2)$ denote the set of integers in the interval $[-q/2, q/2)$.
We define $R_q \coloneqq \bbZ_q[X] / (X^N + 1)$ as the polynomial ring with coefficients in $\bbZ_q$.
For an integer $x \in \bbZ$, the reduction of $x$ modulo $q$ is defined as $x \bmod q \coloneqq x - \floor{ \frac{x + q / 2}{q} } q$, where $\floor{\cdot}$ is the floor function.
For a polynomial $\bmx \in R$ and a vector $x \in \bbZ^n$, this modular reduction is applied to each of their coefficients and elements, respectively.

\subsection{Syntax and homomorphic operations}

RLWE-based encryption schemes consist of the following (probabilistic) polynomial-time algorithms:
\begin{itemize}
    \item
    The setup algorithm $\crs \gets \Setup(1^\lambda)$ takes the unary representation of a security parameter $\lambda \in \bbN$ as input and outputs a common reference string $\crs = (N, q, t)$, where $N, q, t \in \bbN$.
    
    \item
    The key generation algorithm $(\pk, \sk) \gets \KeyGen(\crs)$ takes $\crs$ as input and outputs a public key $\pk$ and a secret key $\sk$.

    \item
    The encryption algorithm $\bmc \gets \Enc(\crs, \pk, \bmm)$ takes $\crs$, $\pk$, and a plaintext $\bmm \in R_t$ as input and outputs a ciphertext $\bmc \in R_q$.

    \item
    The decryption algorithm $\bmm \gets \Dec(\crs, \sk, \bmc)$ takes $\crs$, $\sk$, and a ciphertext $\bmc \in R_q$ as input and outputs a plaintext $\bmm \in R_t$.
\end{itemize}
In whar follows, we assume that $N$ is a power of two, $q$ is a prime, and $t$ is a prime such that $t \bmod 2N = 1$ and $t < q$.
Additionally, we omit the arguments $\crs$, $\pk$, and $\sk$ from these algorithms for simplicity of notation.

RLWE-based encryption schemes satisfy (approximate) correctness, i.e.,
\[
    \Dec(\Enc(\bmm)) = \bmm + \bme \bmod t
\]
for all $\bmm \in R_t$, where $\bme \in R$ is a small error polynomial.
They also support multiplication by any plaintext $\bmk \in R_t$, such that
\[
    \Dec(\bmk \cdot \Enc(\bmm)) = \bmk \cdot \bmm + \bme_\pMult \bmod t
\]
for some $\bme_\pMult \in R$.
Furthermore, there exist binary operations $\oplus: R_q \times R_q \to R_q$ and $\otimes: R_q \times R_q \to R_q$ such that, for all $\bmm_1, \bmm_2 \in R_t$ and their corresponding ciphertexts $\bmc_1 \gets \Enc(\pk, \bmm_1)$ and $\bmc_2 \gets \Enc(\pk, \bmm_2)$, 
\begin{align*}
    \Dec(\bmc_1 \oplus \bmc_2) &= \bmm_1 + \bmm_2 + \bme_\Add \bmod t, \\
    \Dec(\bmc_1 \otimes \bmc_2) &= \bmm_1 \cdot \bmm_2 + \bme_\Mult \bmod t,
\end{align*}
hold for some $\bme_\Add, \bme_\Mult \in R$.

\subsection{Message packing}

In RLWE-based encryption schemes, multiple data elements can be encoded as a single plaintext using the Number Theoretic Transform.
This encoding process is called \emph{packing} and is denoted by $\bmm = \Pack(x)$, where $x \in \bbZ_t^M$ and $\bmm \in R_t$.
To clarify this process, we refer to the integer vector $x$ as the \emph{cleartext}, and to the encoded polynomial $\bmm$ as the \emph{plaintext}.
Additionally, the inverse map of $\Pack$ is called \emph{unpacking} and is denoted by $\Unpack$, satisfying
\[
    \Unpack(\Pack(x)) = x \bmod t
\]
for all $x \in \bbZ_t^M$.
This implies that
\[
    \Unpack(\Dec(\Enc(\Pack(x)))) = x + \epsilon \bmod t
\]
holds for some error vector $\epsilon \in \bbZ^M$.

The packing and unpacking algorithms enable the SIMD (Single Instruction, Multiple Data) operations,
\begin{align*}
    \Unpack(\Pack(x_1) + \Pack(x_2)) &= x_1 + x_2 \bmod t, \\
    \Unpack(\Pack(x_1) \cdot \Pack(x_2)) &= x_1 \circ x_2 \bmod t,
\end{align*}
where $x_1, x_2 \in \bbZ_t^M$, and $\circ$ denotes the Hadamard product.
Combining the packing and unpacking with the RLWE encryption algorithms, the element-wise computations on cleartexts can be performed over $R_q$.
That is, for all $x_1, x_2 \in \bbZ_t^M$, the following equations hold for some error vectors $\epsilon_\Add, \epsilon_\pMult, \epsilon_\Mult \in \bbZ^M$:
\begin{align*}
    &\Unpack(\Dec(\Enc(\Pack(x_1)) \oplus \Enc(\Pack(x_2)))) \\
    &= x_1 + x_2 + \epsilon_\Add \bmod t, \\
    &\Unpack(\Dec(\Pack(x_1) \cdot \Enc(\Pack(x_2)))) \\
    &= x_1 \circ x_2 + \epsilon_\pMult \bmod t, \\
    &\Unpack(\Dec(\Enc(\Pack(x_1)) \otimes \Enc(\Pack(x_2)))) \\
    &= x_1 \circ x_2 + \epsilon_\Mult \bmod t.
\end{align*}
Moreover, there exists a rotation algorithm $\Rotate: R_q \times \bbN \to R_q$ that cyclically shifts the cleartext slots to the left.
For example, let $\bmm = \Pack([a\ b\ c\ d\ e\ f]^\top)$ and $\bmc \gets \Enc(\bmm)$.
Then, we have
\begin{align*}
    &\Unpack(\Dec(\Rotate(\bmc, 2))) \\
    &= [c\ d\ e\ f\ a\ b]^\top + \epsilon_\Rotate \bmod t
\end{align*}
for some $\epsilon_\Rotate \in \bbZ^M$.

In what follows, we use the CKKS encryption scheme~\cite{Cheon17} as an instance of RLWE-based encryption schemes.
In this scheme, $M = N / 2$, and the norms of the error polynomials and vectors are bounded by some constants.
We refer the reader to~\cite{Cheon17,Cheon18} for the detailed derivations of these constants.

\section{Encrypted visual feedback control}

\subsection{Visual feedback control}

\begin{figure}[t]
    \centering
    \includegraphics[width=\columnwidth]{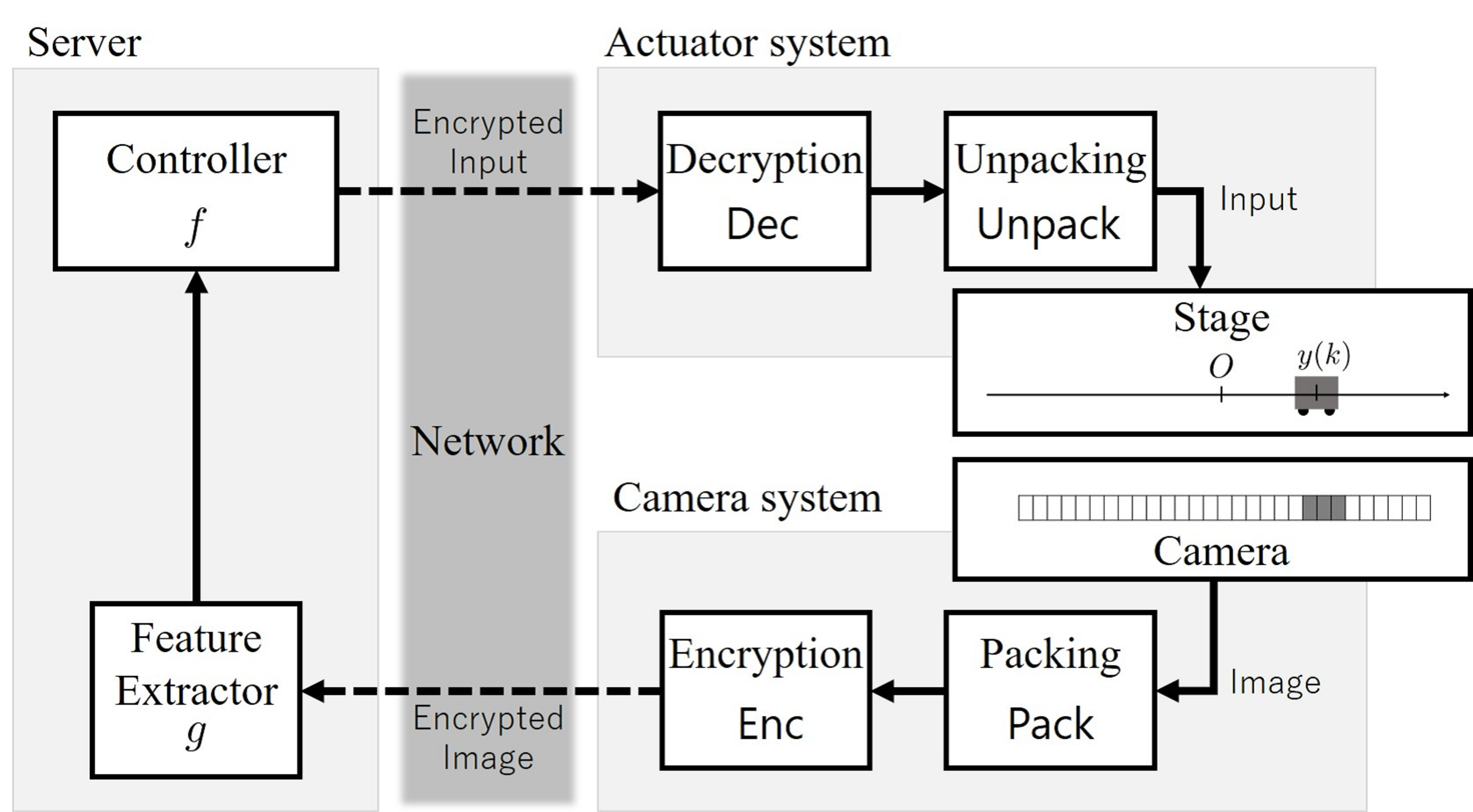}
    \caption{An overview of visual feedback control system.}
    \label{fig:system}
\end{figure}

In this study, we consider the visual feedback control system illustrated in \figref{fig:system}.
The plant and its surrounding environment are monitored by a grayscale camera, which transmits images to a server via a communication network.
The server performs image processing to extract features and computes a control input based on these features.
It then transmits the control input to the actuator, which drives the plant.

Let $k \in \qty{0, 1, 2, \dots}$ be the time index, and let $I(k) \in \qty{0, \dots, 255}^n$ be an image captured by the camera at time $k$.
Specifically, the image consists of $n = n_w n_h$ pixels, each taking an 8-bit brightness value, where $n_w$ and $n_h$ denote the width and height of the image in pixels, respectively.
For each time $k$, the server extracts features from the received image by computing a mapping
\[
    g: \qty{0, \dots, 255}^n \to \bbR^p,
\]
where $p$ is the dimension of the features.

The extracted features $g(I(k))$ are used to compute the control input.
Here, we denote the controller by a mapping
\[
    f: \bbR^p \to \bbR^m.
\]
The control input $u(k)$ at time $k$ is then given by
\begin{equation}
    u(k) \coloneqq f(g(I(k)); \theta),
    \label{eq:input}
\end{equation}
where $\theta$ denotes the controller parameters.

Our goal is to develop an algorithm for evaluating \eqref{eq:input} over encrypted images and controller parameters without decryption.
In the following sections, we describe the specific feature extractor $g$ and controller $f$ considered in this study, and propose an encrypted control scheme for \eqref{eq:input}.

\subsection{Feature extraction}

We focus on the positioning of a one-dimensional stage using visual feedback control.
The stage is driven by the actuator, and its position is captured by the camera.
In this scenario, as it is sufficient to capture its horizontal position, we set $n_h = 1$ in what follows.
Additionally, we assume that the number of pixels, $n = n_w$, is even.

\begin{figure}[t]
    \centering
    \includegraphics[width=\columnwidth]{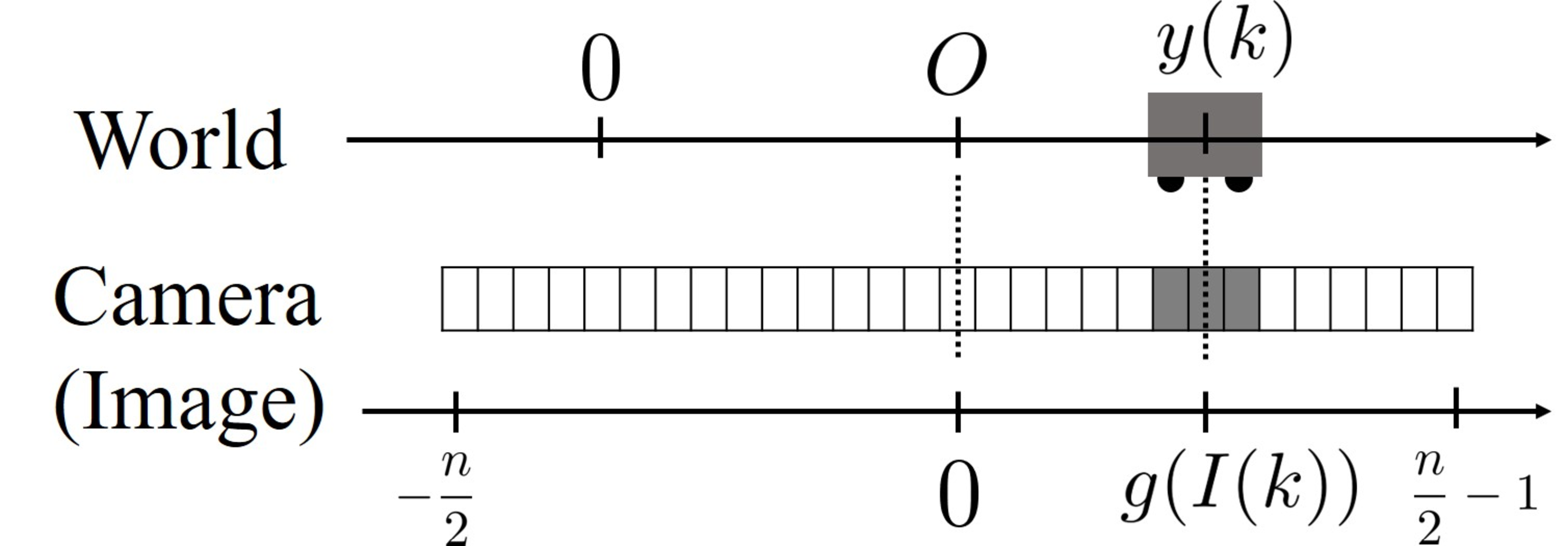}
    \caption{World and camera coordinate systems.}
    \label{fig:coordinates}
\end{figure}

\figref{fig:coordinates} illustrates the world and camera coordinate systems.
In both coordinate systems, the rightward direction is defined as positive.
Let $O$ denote the camera position in the world coordinate system, and assume that the origin of the camera coordinate system coincides with $O$.
Then, in the camera coordinate system, the position of an object in a captured image can be represented by an integer coordinate $i \in \qty{-n / 2, \dots, n / 2 - 1}$.
We regard this discrete value $i$ as a pixel index.

The pixel index representing the stage position is determined by computing the geometric centroid of the stage.
Here, we assume that the stage is sufficiently brighter than its background.
The geometric centroid~\cite{francois02} (i.e., the extracted feature in this scenario) at time $k$ is then given by
\begin{align}
    g(I(k)) &\coloneqq \frac{\calI_w(k)}{\calI(k)}, \label{eq:centroid} \\
    \calI_w(k) &\coloneqq \sum_{i=-n/2}^{n/2-1} i \cdot I_i(k), \label{eq:weighted-brightness} \\
    \calI(k) &\coloneqq \sum_{i=-n/2}^{n/2-1} I_i(k), \label{eq:total-brightness}
\end{align}
where $I_i(k)$ is the brightness value of the $i$-th pixel.

\subsection{Feedback control}

The control objective is to regulate the position of the stage.
Thus, the controller is defined by
\begin{equation}
    f(g(I(k)); K) \coloneqq K g(I(k)),
\label{eq:p-control}
\end{equation}
where the parameter $\theta$ in \eqref{eq:input} corresponds to the feedback gain $K \in \bbR$.
Since the plant is a one-dimensional stage, the control input is a scalar (i.e., $m = 1$).
However, note that it is straightforward to extend the proposed scheme to compute higher-dimensional control inputs.

\subsection{Secure computation using RLWE encryption}

This section proposes the encrypted visual feedback control algorithm for evaluating \eqref{eq:input} with \eqref{eq:centroid} through \eqref{eq:p-control}.
We first observe that the control input \eqref{eq:input} is given by
\[
    u(k) = \frac{K \calI_w(k)}{\calI(k)}.
\]
Our strategy to compute this fraction is (i) to evaluate the numerator and denominator as packed vectors over ciphertexts using homomorphic operations, and (ii) after decryption, to extract their first elements and perform the division.

To formulate this strategy, we define $\rho(\cdot, j)$ as an operator that rotates the elements of a given cleartext to the left by $j$ positions, and define $[ \cdot ]_1$ as an operator that extracts the first element of a given cleartext.
Assume that $M \gg n$ and $t$ is sufficiently large.
Then, for a scaling factor $\Delta > 0$, the scaled numerator and the denominator are respectively approximated as
\begin{align*}
    \Delta K \calI_w(k) &\approx \qty[ x_K \circ \sum_{j=0}^{n-1} \rho(x_w \circ x_I, j) ]_1, \\
    \calI(k) &= \qty[ \sum_{j=0}^{n-1} \rho(x_I, j) ]_1,
\end{align*}
where
\begin{align*}
    x_K &=
    \begin{bmatrix}
        \round{\Delta K} & 0 & \cdots & 0
    \end{bmatrix}^\top
    \bmod t \in \bbZ_t^M, \\
    x_w &=
    \begin{bmatrix}
        -n/2 \!&\! \cdots \!&\! n/2 - 1 \!&\! 0 \!&\! \cdots \!&\! 0
    \end{bmatrix}^\top
    \bmod t \in \bbZ_t^M, \\
    x_I &=
    \begin{bmatrix}
        I^\top(k) & 0 & \cdots & 0
    \end{bmatrix}^\top
    \bmod t \in \bbZ_t^M,
\end{align*}
and $\round{x} \coloneqq \floor{x + 1 / 2}$ denotes the rounding function.
The computations of the right-hand sides can be efficiently performed as
\begin{align*}
    \qty[ x_K \circ \sum_{j=0}^{n-1} \rho(x_w \circ x_I, j) ]_1 &= \qty[ x_K \circ x_I^{w, \ceil{\log_2 n}} ]_1, \\
    \qty[ \sum_{j=0}^{n-1} \rho(x_I, j) ]_1 &= \qty[ x_I^{\ceil{\log_2 n}} ]_1,
\end{align*}
where, for $\ell = 1, \dots, \ceil{ \log_2 n }$,
\begin{align*}
    x_I^{w, \ell} &= x_I^{w, \ell-1} + \rho(x_I^{w, \ell-1}, 2^{\ell-1}), \\
    x_I^\ell &= x_I^{\ell-1} + \rho(x_I^{\ell-1}, 2^{\ell-1}),
\end{align*}
with the initial values $x_I^{w, 0} = x_w \circ x_I$ and $x_I^0 = x_I$, and $\ceil{\cdot}$ is the ceil function.
It should be noted that the feedback gain $K$ is encoded into the cleartext $x_K$ after scaling and rounding\footnote{While the CKKS scheme natively supports encrypting real-valued data and thus does not necessarily require this operation, we incorporate the scaling and rounding process to ensure applicability to general RLWE-based encryption schemes.}.
This conversion is necessary because $K$ is a real number, whereas cleartexts are integer-valued vectors.
The quantization errors induced by this rounding process can be made negligible by choosing a sufficiently large $\Delta$.
Consequently, the control input is computed as
\begin{equation}
    u(k) \approx \Delta^{-1} \frac{ \qty[ x_K \circ x_I^{w, \ceil{\log_2 n}} ]_1 }{ \qty[ x_I^{\ceil{\log_2 n}} ]_1 }.
    \label{eq:approx-input}
\end{equation}

To evaluate the right-hand side of \eqref{eq:approx-input} over ciphertexts, the cleartexts $x_K$, $x_w$, and $x_I$ need to be packed into plaintexts as
\begin{align}
    \bmm_K &= \Pack(x_K), \label{eq:mK} \\
    \bmm_w &= \Pack(x_w), \label{eq:mw} \\
    \bmm_I &= \Pack(x_I). \label{eq:mI}
\end{align}
We then compute the ciphertexts
\begin{align}
    \bmc_K &\gets \Enc(\bmm_K), \label{eq:cK} \\
    \bmc_I &\gets \Enc(\bmm_I). \label{eq:cI}
\end{align}
The ciphertexts corresponding to the numerator and denominator inside the operator $[ \cdot ]_1$ in \eqref{eq:approx-input} are respectively computed by
\begin{align}
    \bmc_\calI^{K,w} &= \bmc_K \otimes \bmc_I^{w, \ceil{\log_2 n}}, \label{eq:ccalIKw} \\
    \bmc_\calI &= \bmc_I^{\ceil{\log_2 n}}, \label{eq:ccalI}
\end{align}
where, for $\ell = 1, \dots, \ceil{\log_2 n}$,
\begin{align}
    \bmc_I^{w, \ell} &= \bmc_I^{w, \ell - 1} \oplus \Rotate(\bmc_I^{w, \ell - 1}, 2^{\ell - 1}), \label{eq:cIwell} \\
    \bmc_I^\ell &= \bmc_I^{\ell - 1} \oplus \Rotate(\bmc_I^{\ell - 1}, 2^{\ell - 1}), \label{eq:cIell}
\end{align}
with the initial values $\bmc_I^{w, 0} = \bmm_w \cdot \bmc_I$ and $\bmc_I^0 = \bmc_I$.
After decryption and unpacking, we obtain
\begin{align*}
    \Unpack(\Dec(\bmc_\calI^{K,w})) &= \qty[\Delta K \calI_w(k)\ \ast\ \cdots \ \ast]^\top + \epsilon_\num, \\
    \Unpack(\Dec(\bmc_\calI)) &= \qty[\calI(k)\ \ast\ \cdots \ \ast]^\top + \epsilon_\den,
\end{align*}
where $\ast$ represents unspecified values, and $\epsilon_\num, \epsilon_\den \in \bbZ^n$ are total errors that include the quantization errors and errors induced by the homomorphic operations.
Therefore, by dividing the first element of $\Unpack(\Dec(\bmc_\calI^{K,w}))$ by that of $\Unpack(\Dec(\bmc_\calI))$ and multiplying by $\Delta^{-1}$, we obtain
\begin{align*}
    \frac{1}{\Delta} \frac{ \qty[ \Unpack(\Dec(\bmc_\calI^{K,w})) ]_1 }{ \qty[ \Unpack(\Dec(\bmc_\calI)) ]_1 } 
    &= \frac{ K \calI_w(k) + \Delta^{-1} [ \epsilon_\num ]_1 }{ \calI(k) + [ \epsilon_\den ]_1 }, \\
    &\approx \frac{ K \calI_w(k) }{ \calI(k) } = u(k),
\end{align*}
for a sufficiently large $\Delta$.

Based on the above formulation, we summarize the proposed algorithm in \algref{alg:evfc}.
In the offline phase, a system designer prepares $\bmm_w$ and $\bmc_K$ using \eqref{eq:mK}, \eqref{eq:mw}, and \eqref{eq:cK}, and transmits them to the server (lines 2--4).
The server then stores the received data.
In the online phase, at each time step $k$, the camera encrypts its captured image $I(k)$ into $\bmc_I$ using \eqref{eq:mI} and \eqref{eq:cI}, and transmits the ciphertext to the server (line 6).
The server computes $\bmc_\calI^{K,w}$ and $\bmc_\calI$ using \eqref{eq:ccalIKw} through \eqref{eq:cIell}, and returns them to the actuator (lines 7--11).
Finally, the actuator recovers the numerator and the denominator, and computes the approximated control input (lines 12--14).

Note that throughout the above derivation, we assume that the cleartexts during computation do not exceed the modulus $t$, thereby avoiding overflow.
In practice, the scaling factor $\Delta$ is constrained by the size of $t$ to prevent this overflow.
The plaintext modulus $t$, in turn, is constrained by the size of the ciphertext modulus $q$, which is ultimately bounded by the desired security level $\lambda$.
Consequently, there exists a fundamental trade-off between the control performance and the security level of the system.
We leave the analysis of this trade-off for future work.

\begin{figure}[t]
    \begin{algorithm}[H]
        \caption{Encrypted visual feedback control}
        \label{alg:evfc}
        \begin{algorithmic}[1]
            \Require $I(k) \in \qty{0, \dots, 255}^n$ and $K \in \bbR$.
            \Ensure $u(k) \in \bbR$.
            \State $\triangleright$ Offline
            \State D: $x_w \gets [-n/2 \ \cdots \ n/2 - 1 \ 0 \cdots \ 0]^\top \bmod t \in \bbZ_t^M$
            \State D: $x_K \gets [\round{\Delta K} \ 0 \ \cdots \ 0]^\top \bmod t \in \bbZ_t^M$
            \State D: $\bmm_w \gets \Pack(x_w)$, $\bmc_K \gets \Enc(\Pack(x_K))$
            \State $\triangleright$ Online
            \State C: $\bmc_I \gets \Enc(\Pack([I^\top(k) \ 0 \ \cdots \ 0]^\top))$
            \State S: $\bmc_I^{w, 0} \gets \bmm_w \cdot \bmc_I$, $\bmc_I^0 \gets \bmc_I$
            \ForAll{$\ell = 1, \dots, \ceil{\log_2 n}$}
                \State S: $\bmc_I^\ell \gets \bmc_I^{\ell - 1} \oplus \Rotate(\bmc_I^{\ell - 1}, 2^{\ell - 1})$
                \State S: $\bmc_I^{w, \ell} \gets \bmc_I^{w, \ell - 1} \oplus \Rotate(\bmc_I^{w, \ell - 1}, 2^{\ell - 1})$
            \EndFor
            \State S: $\bmc_\calI \gets \bmc_I^{\ceil{\log_2 n}}$, $\bmc_\calI^{K,w} \gets \bmc_K \otimes \bmc_I^{w, \ceil{\log_2 n}}$
            \State A: $\den \gets [ \Unpack(\Dec(\bmc_\calI)) ]_1$
            \State A: $\num \gets [ \Unpack(\Dec(\bmc_\calI^{K,w})) ]_1$
            \State A: $u(k) \gets \Delta^{-1} \num / \den$
        \end{algorithmic}
    \end{algorithm}
\end{figure}

\section{Numerical example}

This section verifies the feasibility of \algref{alg:evfc} through numerical simulations.
Suppose that the dynamics of the stage is given by
\[
    y = \frac{ 0.0196 }{ 1 - 0.9804 z^{-1} } u,
\]
where $y$ is the position of the stage, and $u$ is the corresponding control input.
The position of the stage is monitored by a camera at each time step $k$.
To emulate the camera in this simulation, we synthetically generate image data consisting of $n = 500$ pixels, assuming the brightness of the stage and the background to be $10$ and $0$, respectively.
Furthermore, we set the length of the stage within the image to $3$ pixels.
In this setup, we manually tune the feedback gain to $K = 0.8$.

The proposed algorithm was implemented in C++ using the SEAL library~\cite{SEAL} on a computer equipped with an Intel Core i5-9500 CPU and 16~GB of RAM.
The parameters of the CKKS encryption scheme are chosen to achieve $\lambda = 128$-bit security level.
Specifically, we set the polynomial degree to $N = 2^{14}$, and configure the ciphertext modulus $q$ to have a total bit-length of 360~bits\footnote{In the SEAL library, the ciphertext modulus $q$ is formulated as the product of multiple prime numbers. In our implementation, the bit-lengths of these primes are set to $(60, 30, \dots, 30, 60)$, totaling 360 bits.}.
In this simulation, we set $\Delta = 1$ because the CKKS scheme natively supports real-valued cleartexts\footnote{While the RLWE-based schemes in Section~2 use the plaintext modulus $t$, the CKKS scheme employed here does not require it. Instead, the internal scaling factor manages the precision of the real-valued data.}.
However, note that the CKKS scheme internally encodes these real-valued cleartexts into integer representations.
The scaling factor for this internal encoding is set to $2^{30}$.

\begin{figure}[t]
    \centering
    \includegraphics[width=\columnwidth]{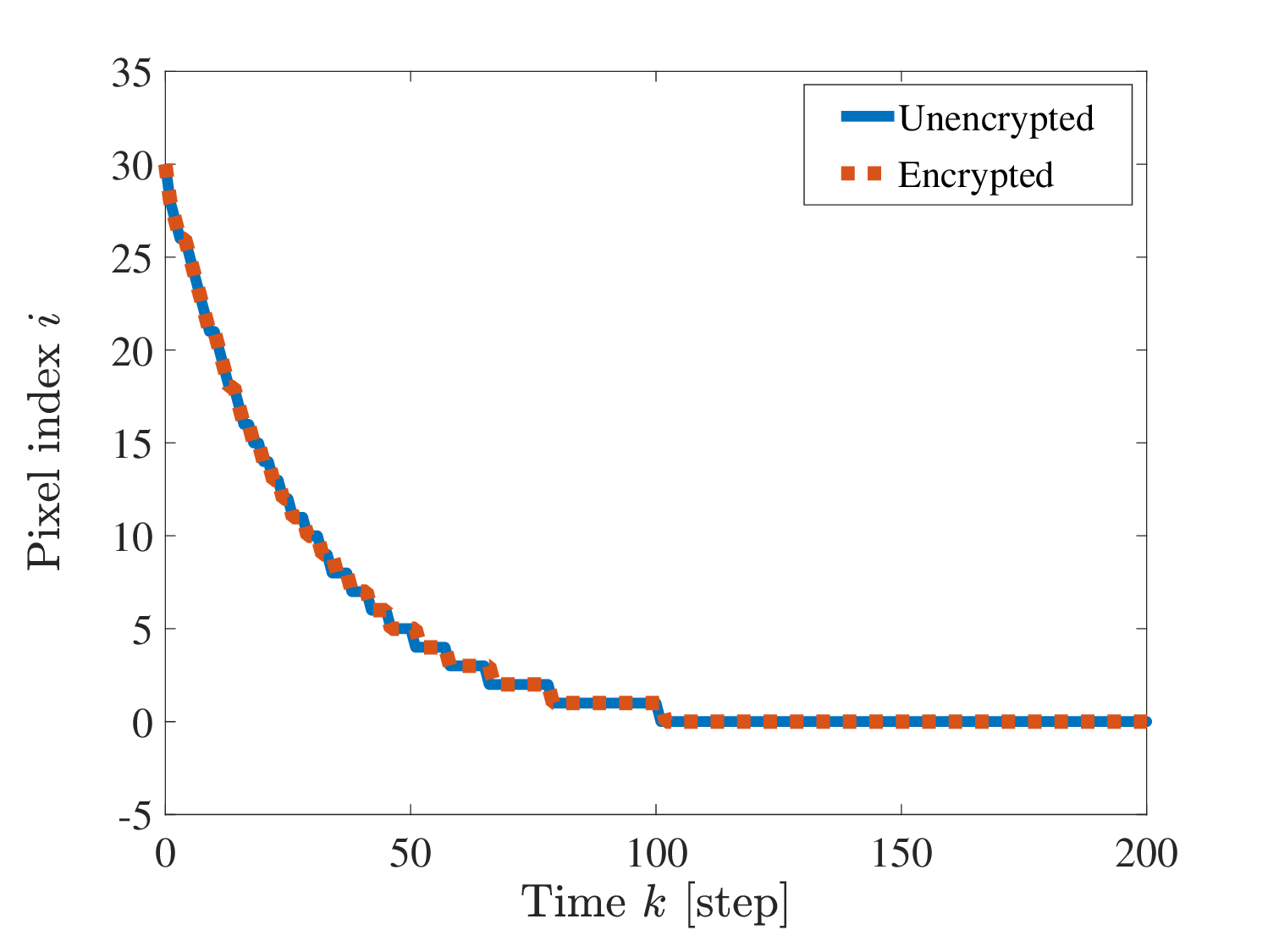}
    \caption{Comparison of the control output (camera view).}
    \label{fig:result}
    \vspace{2em}
\end{figure}

\figref{fig:result} compares the control outputs of the unencrypted visual feedback control and \algref{alg:evfc}, with the initial position of the stage set to $30$.
In this figure, the position of the stage is represented by the pixel index $i \in \{ -255, \dots, 254 \}$.
This result demonstrates that the proposed algorithm achieves a control performance comparable to the original controller, despite the errors introduced by quantization and encryption.

Next, we evaluate the computational cost of \algref{alg:evfc}.
\tabref{tab:time} lists the average computation times required for each component in both the proposed algorithm and a naive implementation without message packing, averaged over $300$ independent trials.
In the naive implementation, each pixel value is regarded as a polynomial of degree zero and encrypted individually, requiring $n$ separate encryption operations.
Moreover, it requires $n$ plaintext-multiplications and $n-1$ additions for computing \eqref{eq:weighted-brightness}, and $n-1$ additions for computing \eqref{eq:total-brightness}.
In contrast, the proposed algorithm significantly reduces the number of encryption operations to just one, thereby achieving a substantial improvement in the computation time at the camera side.
The computation time on the server side is also reduced because the proposed algorithm requires only a single plaintext-multiplication and $\ceil{\log_2 n}$ additions for computing \eqref{eq:weighted-brightness}, and $\ceil{\log_2 n}$ additions for computing \eqref{eq:total-brightness}.
Consequently, the proposed algorithm successfully realizes an efficient and secure visual feedback control by substantially reducing the computational overhead on both the camera and the server.

\begin{table}[t]
    \centering
    \caption{Average computation time (s).}
    \begin{tabular}{@{}clll@{}}
    \hline
                & Camera & Server & Actuator \\
    \hline
    Proposed    & 0.0221 & 0.477  & 0.0128   \\
    w/o packing & 8.89   & 3.71   & 0.0128   \\
    \hline
    \end{tabular}
    \label{tab:time}
\end{table}

\section{Conclusion}

In this study, we proposed an encrypted visual feedback control algorithm for regulating a one-dimensional stage.
The proposed algorithm leverages the message packing technique supported by RLWE encryption schemes to efficiently compute the feature extraction and the control input directly over ciphertexts.
Furthermore, we verified the feasibility of the proposed algorithm through numerical simulations, demonstrating that the errors induced by quantization and encryption can be made negligible by selecting appropriate encryption parameters.
Future work includes analyzing the trade-off between the control performance and the security level achievable by the proposed framework.

\end{document}